\documentclass[aps,prl,twocolumn,superscriptaddress,showpacs,amsmath,amssymb,floatfix]{revtex4}

\usepackage{graphicx}
\usepackage{dcolumn}
\usepackage{bm}

\begin{document}


\title{Quantum Critical Point of Itinerant Antiferromagnet in 
Heavy Fermion}

\author{Hiroaki Kadowaki}
\affiliation{Department of Physics, Tokyo Metropolitan University, 
Hachioji-shi, Tokyo 192-0397, Japan}

\author{Yoshikazu Tabata}
\affiliation{Graduate School of Science, Osaka University, 
Toyonaka, Osaka 560-0043, Japan}

\author{Masugu Sato}
\affiliation{MSD, JASRI, 
1-1-1 Kouto Mikazuki-cho Sayo-gun, Hyogo 679-5198, Japan}

\author{Naofumi Aso}
\affiliation{NSL, Institute for Solid State Physics, 
University of Tokyo, Tokai, Ibaraki 319-1106, Japan}

\author{Stephane Raymond}
\affiliation{CEA-Grenoble, DSM/DRFMC/SPSMS, 38054 Grenoble, France}

\author{Shuzo Kawarazaki}
\affiliation{Graduate School of Science, Osaka University, 
Toyonaka, Osaka 560-0043, Japan}

\date{\today}

\begin{abstract}
A quantum critical point 
of the heavy fermion Ce(Ru$_{1-x}$Rh$_{x}$)$_2$Si$_2$ ($x=0, 0.03$)
has been studied by 
single-crystalline neutron scattering. 
By accurately measuring the dynamical susceptibility 
at the antiferromagnetic wave vector 
$\bm{k}_3 = 0.35 \bm{c}^{\ast}$, 
we have shown that the energy width $\Gamma(\bm{k}_3)$, 
i.e., inverse correlation time, 
depends on temperature as 
$\Gamma(\bm{k}_3) = c_1 + c_2 T^{3/2 \pm 0.1}$, 
where $c_1$ and $c_2$ are $x$ dependent constants, 
in a low temperature range. 
This critical exponent $3/2 \pm 0.1$ 
proves that the quantum critical point is controlled by 
that of the itinerant antiferromagnet.
%
\end{abstract}



\maketitle

Quantum critical points (QCP) separating ferromagnetic or 
antiferromagnetic states from paramagnetic Fermi liquid 
states in strongly correlated electron systems have been 
investigated for decades. 
Successful descriptions of critical behavior 
close to QCPs were traditionally provided by 
the self-consistent renormalization 
(SCR) theory of spin fluctuations 
\cite{MoriyaBook85,Moriya-Ueda03-00} 
for $d$-electron systems based on the Hubbard model. 
The mean-field-type 
approximations used in the SCR theory were justified 
by the renormalization group studies \cite{Hertz76-Millis93} 
based on the Hertz effective action 
above upper critical dimensions. 
For the ferromagnetic QCP, theoretical predictions were 
supported by experimental studies of $d$-electron systems 
\cite{MoriyaBook85,Julian96}. 
In contrast there is little experimental understanding 
of the antiferromagnetic QCP \cite{Moriya-Ueda03-00}. 

A recent intriguing issue of QCP under controversial 
debate is directed toward revealing relevant fixed points for 
antiferromagnetic QCPs in heavy-fermion systems 
\cite{Coleman01}. 
In energy scales much lower than 
Kondo temperature $T_{\text{K}}$, 
$f$ and conduction electrons form composite 
quasiparticles with a large mass renormalization in 
paramagnetic heavy fermions. 
By tuning a certain parameter, e.g., pressure or concentration, 
an antiferromagnetic long range order emerges 
from the Fermi liquid state. 
In a weak coupling picture, 
it has been hypothesized that 
the same QCP as the $d$-electron 
itinerant antiferromagnet, referred to as spin density wave (SDW) 
type QCP, is relevant to the heavy fermion QCP 
\cite{Moriya-Taki,Coleman01}. 

However despite a number of experimental studies 
of heavy-fermion systems showing non-Fermi liquid behavior, 
none of them definitely supports the SDW QCP 
\cite{Lohneysen94,Kambe96,Stewart01}. 
This stems partly from experimental difficulty in measuring 
weakly divergent quantities around QCP 
especially for bulk properties, 
which has been also the case 
for $d$-electron itinerant antiferromagnets \cite{Moriya-Ueda03-00}. 
On the contrary, several recent experiments suggest the 
possibility of a novel strong coupling picture of the 
QCP \cite{Coleman01,Schroder00,Paschen04}.
Among these studies direct measurements of the diverging 
spin fluctuation using single-crystalline neutron scattering 
for the heavy fermion CeCu$_{5.9}$Au$_{0.1}$ 
provided interesting insight \cite{Schroder00}. 
On the basis of the observed 
$E/T$ scaling with an anomalous exponent \cite{Schroder00} 
and effective two space dimensions \cite{Stockert98}, 
a scenario of a locally critical 
QCP was proposed \cite{Coleman01,Si01}. 
In contrast to the SDW QCP, this theory stresses that 
separation of $f$ spin from the quasiparticle state 
occurs abruptly at the QCP. 

\begin{figure}
\begin{center}
\includegraphics[width=8.5cm,clip]{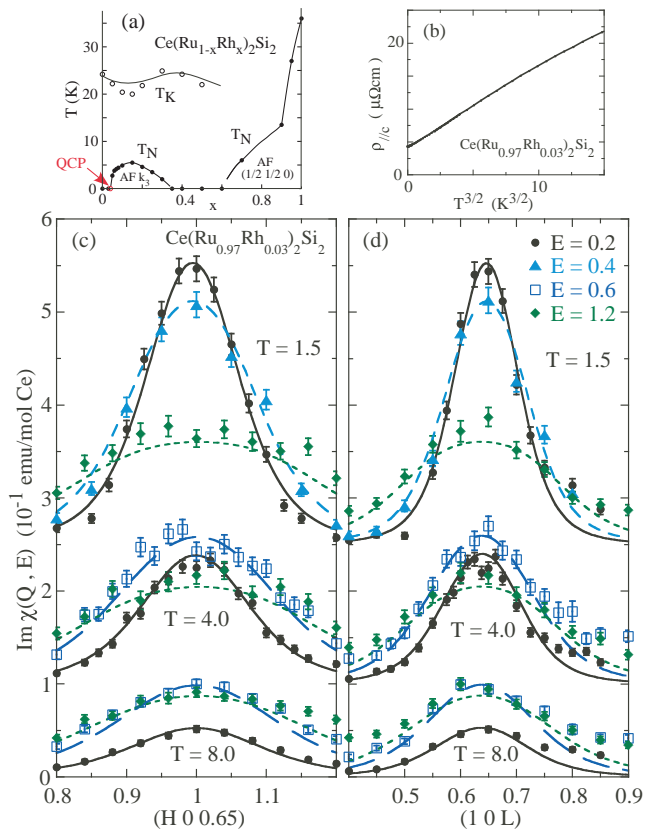}
\end{center}
\caption{
\label{fig:Qscan}
(color online) (a) The phase diagram and $T_{\text{K}}$ 
of Ce(Ru$_{1-x}$Rh$_{x}$)$_2$Si$_2$ 
are reproduced from Refs.~\cite{Sekine92,Kawarazaki97}. 
(b) Resistivity of the sample with $x = 0.03$ is plotted 
as a function of $T^{3/2}$. 
Constant-$E$ scans taken with $E = 0.2$, 0.4, 0.6 and 1.2 meV 
along (c) $\bm{Q} = (H,0,0.65)$ and (d) $(1,0,L)$ lines 
for the sample with $x = 0.03$. 
Data of $T = 1.5$ and 4 K are shifted by 
0.25 and 0.1 emu/mol Ce, respectively for clarity.
Curves in (c) and (d) are calculations using 
Eqs.~(\ref{eq:ImChi1}) and (\ref{eq:ImChi2}), 
corrected for resolution functions with the same fit parameter as 
those shown in Fig.~\ref{fig:Escan}. 
}
\end{figure}

In the present work, we have studied an antiferromagnetic 
QCP of another heavy-fermion 
system Ce(Ru$_{1-x}$Rh$_{x}$)$_2$Si$_2$ ($x=0$, $0.03$) 
using single-crystalline neutron scattering. 
Stoichiometric CeRu$_2$Si$_2$ is an archetypal paramagnetic 
heavy-fermion with enhanced $C/T \simeq 350$ mJ/K$^2$ mol 
and $T_{\text{K}} \simeq 24$ K \cite{Besnus85}. 
Extensive neutron scattering studies of 
CeRu$_2$Si$_2$ \cite{Kadowaki04} 
have shown that spin fluctuations possessing 
three-dimensional ($d=3$) character are excellently 
described by the SCR theory for heavy fermions \cite{Moriya-Taki}. 
A small amount of Rh doping 
$x > x_{\text{c}} \simeq 0.04$ \cite{Sekine92} 
induces an antiferromagnetic phase 
[see Fig.~\ref{fig:Qscan}(a)] 
of the sinusoidally modulated structure 
with the wave vector $\bm{k}_3 = 0.35 \bm{c}^{\ast}$ 
\cite{Kawarazaki97}. 

The Rh doping modifies exchange interactions, 
while keeping $T_{\text{K}}$ constant \cite{Sekine92}, 
and brings about the antiferromagnetic phase in 
the concentration range $x_{\text{c}} < x < 0.35$. 
The dome structure of this phase and 
another antiferromagnetic phase with the modulation 
vector $(\frac{1}{2} \frac{1}{2} 0)$ 
in the higher concentration range $0.6 < x$ suggests 
a certain frustration effect among exchange interactions. 
Samples nearly tuned to the lowest concentration 
QCP ($x \sim x_{\text{c}}$) show non-divergent 
$C/T$ ($T \rightarrow 0$) \cite{Tabata98} and 
$\Delta \rho \propto T^{3/2}$ 
[see Fig.~\ref{fig:Qscan}(b)], 
which are consistent with the SDW QCP in $d=3$. 
Thus one can expect that Ce(Ru$_{1-x}$Rh$_{x}$)$_2$Si$_2$ 
($x \lesssim x_{\text{c}}$) is suited to 
investigate the SDW QCP without 
disorder effects. 
On the other hand, non-Fermi liquid 
behavior observed in the higher concentration range 
$0.35 < x$ was reported to be influenced by disorders 
\cite{Taniguchi99,Tabata01}. 

In order to experimentally show the SDW QCP in $d=3$, 
spin correlation studied by the renormalization group theory 
\cite{Hertz76-Millis93,SachdevBook99} 
should be measured directly 
by neutron scattering. 
The theory of the SDW QCP shows that the 
wave-vector dependent susceptibility for the tuned 
sample ($x = x_{\text{c}}$) diverges 
as $\chi(\bm{k}_3) \propto T^{-3/2}$ 
\cite{MoriyaBook85,Hertz76-Millis93}, 
or the characteristic energy of the spin fluctuation, 
i.e., the inverse correlation time, 
depends on temperature as 
$\Gamma(\bm{k}_3) \propto \chi(\bm{k}_3)^{-1} \propto T^{3/2}$. 
By taking the detuning effect ($x < x_{\text{c}}$) into account, 
leading two terms of $\Gamma(\bm{k}_3)$ 
computed by the renormalization group theory 
\cite{Hertz76-Millis93,SachdevBook99} 
are given by 
\begin{equation}
\label{eq:T1.5}
\Gamma(\bm{k}_3) 
=
 c_1 + c_2 T^{3/2} 
\; ,
\end{equation}
where $c_{1}$ and $c_{2}$ are $x$ dependent constants. 
This equation is an approximation in the temperature range 
$T_{\text{FL}} \ll T \ll T_{\text{K}}$, 
where $T_{\text{FL}}$ is a crossover temperature below which 
system shows the Fermi liquid behavior 
\cite{Hertz76-Millis93,SachdevBook99}. 
In the present work, we have accurately measured $\Gamma(\bm{k}_3)$ 
and have shown that it agrees well with 
Eq.~(\ref{eq:T1.5}) 
for both the nearly tuned sample $x=0.03$ and 
the stoichiometric sample $x=0$, which indicates that 
disorder does not influence the critical behavior. 

Neutron-scattering measurements were performed on the triple-axis 
spectrometer HER at the Japan Atomic Energy Research Institute. 
It was operated using final energies of 
$E_{\text{f}}=3.1$ and $2.4$ meV providing energy resolutions 
of $0.1$ and $0.05$ meV 
(full width at half maximum), respectively, at elastic positions. 
Single crystals with a total weight of 19 g ($x=0$) and 
17 g ($x=0.03$) were grown by the 
Czochralski method. 
Two sets of multicrystal samples aligned together 
were mounted in a He flow cryostat so as to measure 
a $(h0l)$ scattering plane. 
All the data shown are converted to the dynamical susceptibility. 
It is scaled to absolute units by comparison with the intensity of 
a standard vanadium sample. 
We note that a new point of the present work is unprecedented 
experimental accuracy in determining the critical 
exponent [3/2 in Eq.~(\ref{eq:T1.5})] using large samples 
and long counting time. 
This has enabled us to determine the singularity 
of QCP and to make qualitative conclusion of the universality class. 
In the pioneering work using the related compound 
Ce$_{1-x}$La$_{x}$Ru$_2$Si$_2$ \cite{Kambe96,Raymond01}, 
Eq.~(\ref{eq:T1.5}) has been discussed assuming 
the fixed value of the exponent, which could be 
determined only with an order larger experimental error. 

The imaginary part of the dynamical susceptibility 
at $\bm{Q}=\bm{k}_{3}+\bm{q}$ 
with small $|\bm{q}|$ and $|E|$
is predicted to be approximated by 
\begin{subequations}
\label{eq:ImChi}
\begin{eqnarray}
\text{Im}\chi(\bm{k}_{3}+\bm{q}, E) 
&=&
\frac{ \chi(\bm{k}_{3}) \Gamma(\bm{k}_{3}) E }
{ E^2 + \Gamma(\bm{k}_{3}+\bm{q})^2 }
\; ,\label{eq:ImChi1}
\\
\Gamma(\bm{k}_{3}+\bm{q}) 
&=& 
D_c [\kappa_c^2 + q_c^2 + F (q_a^2 + q_b^2)]
\; ,\label{eq:ImChi2}
\end{eqnarray}
\end{subequations}
where $D_c$ and $F$ are $T$ independent parameters, and 
$\kappa_c$ is inverse correlation length along the $c$ axis 
\cite{expansionQ}. 
This expansion form with the $T$ independent 
product $\chi (\bm{k}_{3}) \Gamma (\bm{k}_{3})$ has been used 
in the SCR 
\cite{Moriya-Taki,Moriya-Ueda03-00} and 
renormalization group \cite{SachdevBook99} theories. 
The two parameters $D_c$ and $F$ were determined by 
using constant-$Q$ and \mbox{-$E$} scans for both samples with 
$x=0$ \cite{Kadowaki04} and $0.03$ at $T=1.5$ K. 
These data were fitted to Eqs.~(\ref{eq:ImChi1}) 
and (\ref{eq:ImChi2}) 
convolved with the resolution functions. 
In Figs.~\ref{fig:Qscan}(c) and \ref{fig:Qscan}(d) 
we show constant-$E$ scans 
through the antiferromagnetic wave vector $\bm{Q}=(101)-\bm{k}_{3}$ 
and the fit curves for the sample with $x=0.03$. 
The good quality of fitting indicates that 
Eqs.~(\ref{eq:ImChi1}) and (\ref{eq:ImChi2}) well 
describe the experimental data 
at $T=1.5$ K. 
We obtained concentration independent values of the 
parameters $D_c = 98 \pm 4$ (meV \AA$^2$) and 
$F = 0.12 \pm 0.01$. 

\begin{figure}
\begin{center}
\includegraphics[width=7.8cm,clip]{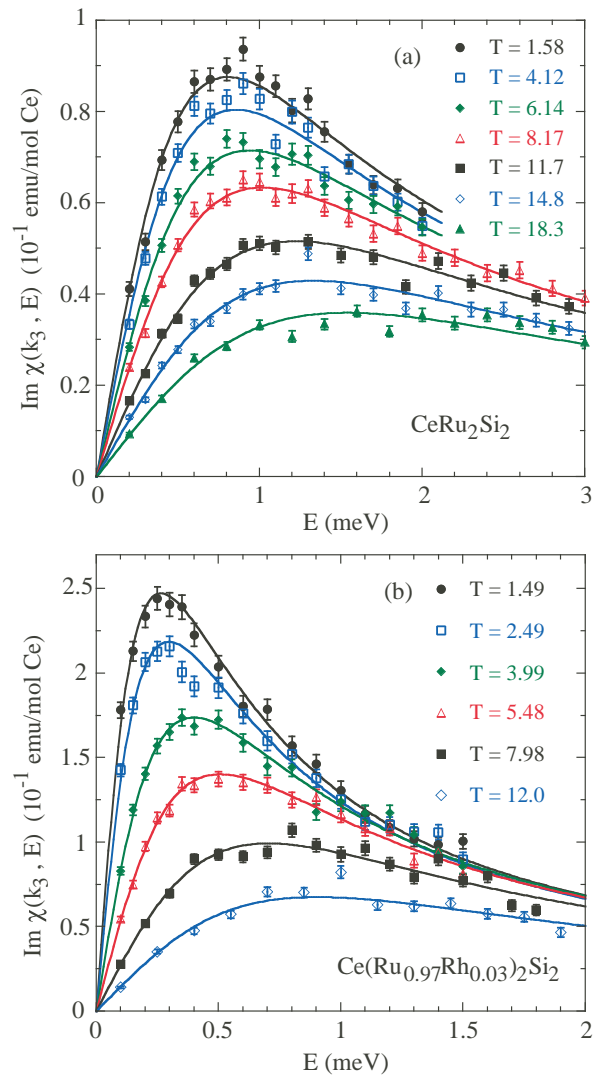}
\end{center}
\caption{
\label{fig:Escan}
(color online) Constant-$Q$ scans 
measured at the antiferromagnetic wave vector 
$\bm{Q}=(101)-\bm{k}_{3}$ 
for the samples with $x=0$ (a) and $x=0.03$ (b).
Curves are fit results using Eq.~(\ref{eq:ImChi1}) with 
two adjustable parameters 
$\Gamma (\bm{k}_{3})$ and $\chi (\bm{k}_{3})$. 
}
\end{figure}

Temperature dependence of 
$\Gamma (\bm{k}_{3}) = D_{c} \kappa_{c}^{2}$ 
has been determined 
by performing constant-$Q$ scans taken at 
$\bm{Q}=(101)-\bm{k}_{3}$ with much 
higher statistical accuracy 
than previous measurements \cite{Raymond01,Kadowaki04}. 
These scan data were fitted to Eq.~(\ref{eq:ImChi1}) 
convolved with the resolution functions, 
where there are two adjustable parameters 
$\Gamma (\bm{k}_{3})$ and $\chi (\bm{k}_{3})$. 
Several fit results of the constant-$Q$ scans 
for the samples with $x=0$ and 0.03 are 
shown in Fig.~\ref{fig:Escan}. 
From these figures one can see that the quality of fitting 
is excellent. 
We also checked the $T$ independence of the 
parameters $D_c$ and $F$ 
by comparing the constant-$E$ scans in Fig.~\ref{fig:Qscan} 
at $T=4$ and 8 K with those calculated using the $T$ dependent 
$\Gamma (\bm{k}_{3})$ and $\chi (\bm{k}_{3})$ determined 
by the constant-$Q$ scans. 
The calculated curves in Fig.~\ref{fig:Qscan}, which 
have no adjustable parameters for $T=4$ and 8 K, 
agree reasonably 
well with the observations. 
Thus we conclude that the theoretical approximation of 
Eqs.~(\ref{eq:ImChi}) has been experimentally confirmed, 
and that the fit parameter $\Gamma (\bm{k}_{3})$ has been 
determined very precisely. 

\begin{figure}
\begin{center}
\includegraphics[width=8.4cm,clip]{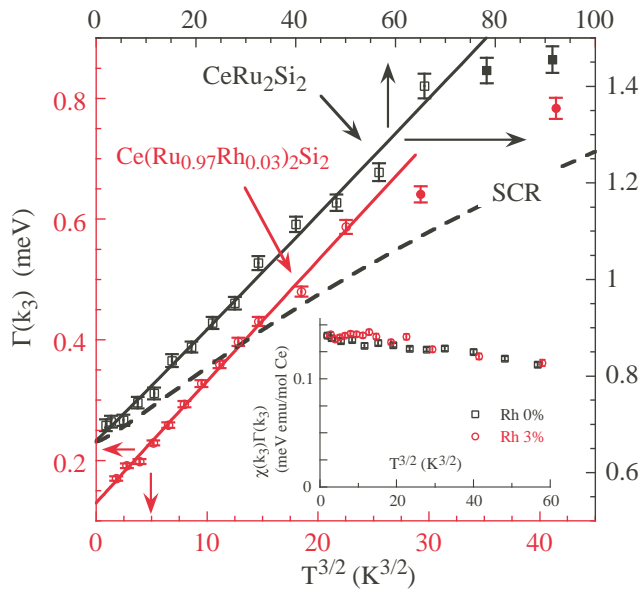}
\end{center}
\caption{
\label{fig:Gammak3}
(color online) 
Energy width $\Gamma (\bm{k}_{3})$ 
of the Lorentzian form Eq.~(\ref{eq:ImChi1}) 
is plotted as a function of $T^{3/2}$. 
Full lines are fit to $\Gamma (\bm{k}_{3}) = c_1 + c_2 T^{v}$ 
with adjustable parameters $c_1$, $c_2$, and $v$ 
in low temperature ranges, 
where data are displayed by open symbols. 
The dashed line is the calculation \cite{Kadowaki04} 
using the SCR theory \cite{Moriya-Taki}. 
The inset shows temperature dependence of 
the product $\chi(\bm{k}_{3}) \Gamma(\bm{k}_{3})$.
}
\end{figure}

Temperature dependence of $\Gamma (\bm{k}_{3})$ is 
shown in Fig.~\ref{fig:Gammak3} by plotting data 
as a function of $T^{3/2}$. 
At low temperatures 
observed data clearly agree with 
the linear behavior of 
Eq.~(\ref{eq:T1.5}). 
In fact, by least squares fitting we obtained 
$\Gamma (\bm{k}_{3}) = (0.67 \pm 0.01) 
+ (0.0095 \pm 0.0021) T^{1.53 \pm 0.08}$ 
(in units of meV) in 
$1.5<T<16$ K for the sample with $x=0$ and 
$\Gamma (\bm{k}_{3}) = (0.129 \pm 0.007) 
+ (0.020 \pm 0.003) T^{1.49 \pm 0.07}$ 
in $1.5<T<8$ K for $x=0.03$. 
Therefore we conclude that the observed critical 
exponent $3/2 \pm 0.1$ is in agreement with the 
theoretical value $3/2$, and consequently that the 
temperature dependence of the spin fluctuation 
is controlled by the SDW QCP in $d=3$. 
The same exponent for both $x=0$ and 0.03 samples 
ensures that the randomness due to Rh doping does not 
affect the criticality.
The temperature independence of 
$\chi (\bm{k}_{3}) \Gamma (\bm{k}_{3})$ 
was also confirmed as shown 
in the inset of Fig.~\ref{fig:Gammak3}. 

The constant $c_1$ in Eq.~(\ref{eq:T1.5}) is proportional to 
the theoretical tuning parameter, a coefficient of the 
quadratic terms of the Hertz effective action, 
and $c_{1} \propto x_{\text{c}}-x$ is normally assumed 
\cite{Hertz76-Millis93}. 
This assumption is consistent with the observed values of 
$c_1$ and the critical concentration 
$x_{\text{c}}=0.04 \pm 0.005$. 
In contrast to this agreement, 
the concentration dependence of $c_2$, 
$c_2(x=0.03)/c_2(x=0) \sim 2$, may suggest a certain 
difficulty in the theoretical interpretation. 
The constant $c_2$, being proportional to the coefficient 
of the quartic term of the Hertz effective action, is usually supposed 
to vary slowly in the concentration range of interest 
\cite{Hertz76-Millis93}. 
The appreciable variation of $c_2$ seems to suggest 
unknown perturbations for 
Ce(Ru$_{1-x}$Rh$_{x}$)$_2$Si$_2$. 
In terms of the SCR theories, the variation of $c_2$ may be 
accounted for by the adjustable mode-mode coupling constant 
($\propto c_2$) that is used in the phenomenological SCR theory 
developed for $d$-electron systems 
\cite{MoriyaBook85,Moriya-Ueda03-00}. 
Despite this problem, we think that the critical exponent $3/2$ 
determined by basic characteristics of the 
system, the space dimension $d=3$ and the dynamical 
exponent $z=2$, 
is more important and decisive to conclude 
the nature of the QCP.

An advantage of the present neutron scattering study is that 
Eq.~(\ref{eq:T1.5}) holds 
in a wider temperature range 
compared to those of indirect measurements using bulk 
properties, 
e.g., $C/T = \gamma_0 - \alpha T^{1/2}$ or 
$\Delta \rho \propto T^{3/2}$ \cite{Kambe96,Tabata98}. 
Theoretically Eq.~(\ref{eq:T1.5}) is an approximation in the 
temperature range $T_{\text{FL}} \ll T \ll T_{\text{K}}$, 
where $T_{\text{FL}}$ is the crossover temperature 
to the Fermi liquid state 
\cite{Hertz76-Millis93,SachdevBook99}. 
The temperature range in which Eq.~(\ref{eq:T1.5}) 
is observable can be discussed quantitatively 
using the SCR theory \cite{Moriya-Taki}. 
In Fig.~\ref{fig:Gammak3}, the dashed line reproduces 
the SCR computation of $\Gamma (\bm{k}_{3})$ 
for CeRu$_2$Si$_2$ based on the previous neutron 
scattering study \cite{Kadowaki04}. 
Apart from discrepancy of the coefficient $c_2$, one can see that 
the $T^{3/2}$ dependence of Eq.~(\ref{eq:T1.5}) 
is a good approximation for the SCR curve in the 
$T$ range $2.5 < T < 13$ K ($4<T^{3/2}<47$), which agrees 
with that of the observed data for CeRu$_2$Si$_2$. 
Since the lower bound temperature is shown to be 
proportional to the tuning parameter by 
the renormalization group theory 
\cite{Hertz76-Millis93,SachdevBook99}, 
the $T^{3/2}$ dependence can be expected 
in a $T$ range of $0.5$ K $< T$ 
($0.35 < T^{3/2}$) for the sample with $x=0.03$. 
The smaller $T$ range of the $T^{3/2}$ dependence for $x=0.03$ 
is probably related to the larger constant $c_2(x=0.03)$. 
We note that below 2.5 K 
the SCR computation of $\Gamma (\bm{k}_{3})$ for 
$x=0$ \cite{Kadowaki04} is approximated by the Fermi liquid 
behavior of 
$\Gamma (\bm{k}_{3},T)-\Gamma (\bm{k}_{3},0) \propto T^2$ 
\cite{SachdevBook99}, 
which is not clearly seen within the present experimental error. 

In connection with neutron scattering experiments of 
CeCu$_{5.9}$Au$_{0.1}$ \cite{Schroder00,Stockert98}, 
it was theoretically predicted ~\cite{Si01} 
that the locally critical QCP is relevant for the 
two-dimensional spin fluctuation, 
in agreement with the experiments of CeCu$_{5.9}$Au$_{0.1}$. 
This theory also predicted that the SDW QCP is relevant for 
the three-dimensional spin fluctuation, which is 
in accord with the present results. 
Finally we note that the present work is first clear experimental 
verification of the SDW QCP to our knowledge 
among single-crystalline neutron scattering 
studies on QCP or non-Fermi liquid behavior of heavy fermions, e.g., 
Refs.~\cite{Schroder00,Stockert98,Raymond01,Kadowaki03,Knafo04,Fak05} 
and $d$-electron systems, e.g., 
Refs.~\cite{Aeppli97,Bao98}. 
Assuming that criticalities of QCPs are classified into 
a limited number of universality classes, we expect that 
the SDW QCP remains to be observed in other systems.

In summary, we have demonstrated that the quantum critical 
behavior of the heavy fermion Ce(Ru$_{1-x}$Rh$_{x}$)$_2$Si$_2$ 
is controlled by the SDW type QCP in three space dimensions. 
The inverse correlation time, i.e., energy width 
$\Gamma(\bm{k}_3)$ of the dynamical susceptibility, 
shows the $T^{3/2}$ dependence predicted by 
the renormalization group and SCR theories.

We wish to acknowledge B. F{\aa}k, J. Flouquet, T. Taniguchi, 
and T. Moriya for valuable discussions.


\end{document}